\documentclass[journal]{IEEEtran}
\usepackage[dvips]{graphicx,color}
\usepackage{latexsym}
\usepackage{amssymb}
\usepackage{amsmath,bm}
\usepackage{array}

\begin{document}

%
%
\newcommand{\qed}{\hfill$\square$}
\newcommand{\suchthat}{\mbox{~s.t.~}}
\newcommand{\markov}{\leftrightarrow}
\newenvironment{pRoof}{%
 \noindent{\em Proof.\ }}{%
 \hspace*{\fill}\qed \\
 \vspace{2ex}}


\newcommand{\ket}[1]{| #1 \rangle}
\newcommand{\bra}[1]{\langle #1 |}
\newcommand{\bol}[1]{\mathbf{#1}}
\newcommand{\rom}[1]{\mathrm{#1}}
\newcommand{\san}[1]{\mathsf{#1}}
\newcommand{\mymid}{:~}
\newcommand{\argmax}{\mathop{\rm argmax}\limits}
\newcommand{\argmin}{\mathop{\rm argmin}\limits}

\newcommand{\Cls}{class NL}
\newcommand{\vSpa}{\vspace{0.3mm}}
\newcommand{\Prmt}{\zeta}
\newcommand{\pj}{\omega_n}

\newfont{\bg}{cmr10 scaled \magstep4}
\newcommand{\bigzerol}{\smash{\hbox{\bg 0}}}
\newcommand{\bigzerou}{\smash{\lower1.7ex\hbox{\bg 0}}}
\newcommand{\nbn}{\frac{1}{n}}
\newcommand{\ra}{\rightarrow}
\newcommand{\la}{\leftarrow}
\newcommand{\ldo}{\ldots}
\newcommand{\typi}{A_{\epsilon }^{n}}
\newcommand{\bx}{\hspace*{\fill}$\Box$}
\newcommand{\pa}{\vert}
\newcommand{\ignore}[1]{}

%
%
%
%
\newcommand{\bc}{\begin{center}}  %
\newcommand{\ec}{\end{center}}
\newcommand{\befi}{\begin{figure}[h]}  %
\newcommand{\enfi}{\end{figure}}
\newcommand{\bsb}{\begin{shadebox}\begin{center}}   %
\newcommand{\esb}{\end{center}\end{shadebox}}
\newcommand{\bs}{\begin{screen}}     %
\newcommand{\es}{\end{screen}}
\newcommand{\bib}{\begin{itembox}}   %
\newcommand{\eib}{\end{itembox}}
\newcommand{\bit}{\begin{itemize}}   %
\newcommand{\eit}{\end{itemize}}
\newcommand{\defeq}{\stackrel{\triangle}{=}}
\newcommand{\Qed}{\hbox{\rule[-2pt]{3pt}{6pt}}}
\newcommand{\beq}{\begin{equation}}
\newcommand{\eeq}{\end{equation}}
\newcommand{\beqa}{\begin{eqnarray}}
\newcommand{\eeqa}{\end{eqnarray}}
\newcommand{\beqno}{\begin{eqnarray*}}
\newcommand{\eeqno}{\end{eqnarray*}}
\newcommand{\ba}{\begin{array}}
\newcommand{\ea}{\end{array}}
\newcommand{\vc}[1]{\mbox{\boldmath $#1$}}
\newcommand{\lvc}[1]{\mbox{\scriptsize \boldmath $#1$}}
\newcommand{\svc}[1]{\mbox{\scriptsize\boldmath $#1$}}

\newcommand{\wh}{\widehat}
\newcommand{\wt}{\widetilde}
\newcommand{\ts}{\textstyle}
\newcommand{\ds}{\displaystyle}
\newcommand{\scs}{\scriptstyle}
\newcommand{\vep}{\varepsilon}
\newcommand{\rhp}{\rightharpoonup}
\newcommand{\cl}{\circ\!\!\!\!\!-}
\newcommand{\bcs}{\dot{\,}.\dot{\,}}
\newcommand{\eqv}{\Leftrightarrow}
\newcommand{\leqv}{\Longleftrightarrow}
\newtheorem{co}{Corollary} 
\newtheorem{lm}{Lemma} 
\newtheorem{Ex}{Example} 
\newtheorem{Th}{Theorem}
\newtheorem{df}{Definition} 
\newtheorem{pr}{Property} 
\newtheorem{pro}{Proposition} 
\newtheorem{rem}{Remark} 

\newcommand{\lcv}{convex } 

\newcommand{\hugel}{{\arraycolsep 0mm
                    \left\{\ba{l}{\,}\\{\,}\ea\right.\!\!}}
\newcommand{\Hugel}{{\arraycolsep 0mm
                    \left\{\ba{l}{\,}\\{\,}\\{\,}\ea\right.\!\!}}
\newcommand{\HUgel}{{\arraycolsep 0mm
                    \left\{\ba{l}{\,}\\{\,}\\{\,}\vspace{-1mm}
                    \\{\,}\ea\right.\!\!}}
\newcommand{\huger}{{\arraycolsep 0mm
                    \left.\ba{l}{\,}\\{\,}\ea\!\!\right\}}}

\newcommand{\Huger}{{\arraycolsep 0mm
                    \left.\ba{l}{\,}\\{\,}\\{\,}\ea\!\!\right\}}}

\newcommand{\HUger}{{\arraycolsep 0mm
                    \left.\ba{l}{\,}\\{\,}\\{\,}\vspace{-1mm}
                    \\{\,}\ea\!\!\right\}}}

\newcommand{\hugebl}{{\arraycolsep 0mm
                    \left[\ba{l}{\,}\\{\,}\ea\right.\!\!}}
\newcommand{\Hugebl}{{\arraycolsep 0mm
                    \left[\ba{l}{\,}\\{\,}\\{\,}\ea\right.\!\!}}
\newcommand{\HUgebl}{{\arraycolsep 0mm
                    \left[\ba{l}{\,}\\{\,}\\{\,}\vspace{-1mm}
                    \\{\,}\ea\right.\!\!}}
\newcommand{\hugebr}{{\arraycolsep 0mm
                    \left.\ba{l}{\,}\\{\,}\ea\!\!\right]}}
\newcommand{\Hugebr}{{\arraycolsep 0mm
                    \left.\ba{l}{\,}\\{\,}\\{\,}\ea\!\!\right]}}

\newcommand{\HugebrB}{{\arraycolsep 0mm
                    \left.\ba{l}{\,}\\{\,}\vspace*{-1mm}\\{\,}\ea\!\!\right]}}

\newcommand{\HUgebr}{{\arraycolsep 0mm
                    \left.\ba{l}{\,}\\{\,}\\{\,}\vspace{-1mm}
                    \\{\,}\ea\!\!\right]}}

\newcommand{\hugecl}{{\arraycolsep 0mm
                    \left(\ba{l}{\,}\\{\,}\ea\right.\!\!}}
\newcommand{\Hugecl}{{\arraycolsep 0mm
                    \left(\ba{l}{\,}\\{\,}\\{\,}\ea\right.\!\!}}
\newcommand{\hugecr}{{\arraycolsep 0mm
                    \left.\ba{l}{\,}\\{\,}\ea\!\!\right)}}
\newcommand{\Hugecr}{{\arraycolsep 0mm
                    \left.\ba{l}{\,}\\{\,}\\{\,}\ea\!\!\right)}}

\newcommand{\hugepl}{{\arraycolsep 0mm
                    \left|\ba{l}{\,}\\{\,}\ea\right.\!\!}}
\newcommand{\Hugepl}{{\arraycolsep 0mm
                    \left|\ba{l}{\,}\\{\,}\\{\,}\ea\right.\!\!}}
\newcommand{\hugepr}{{\arraycolsep 0mm
                    \left.\ba{l}{\,}\\{\,}\ea\!\!\right|}}
\newcommand{\Hugepr}{{\arraycolsep 0mm
                    \left.\ba{l}{\,}\\{\,}\\{\,}\ea\!\!\right|}}

\newcommand{\MEq}[1]{\stackrel{
{\rm (#1)}}{=}}

\newcommand{\MLeq}[1]{\stackrel{
{\rm (#1)}}{\leq}}

\newcommand{\ML}[1]{\stackrel{
{\rm (#1)}}{<}}

\newcommand{\MGeq}[1]{\stackrel{
{\rm (#1)}}{\geq}}

\newcommand{\MG}[1]{\stackrel{
{\rm (#1)}}{>}}

\newcommand{\MPreq}[1]{\stackrel{
{\rm (#1)}}{\preceq}}

\newcommand{\MSueq}[1]{\stackrel{
{\rm (#1)}}{\succeq}}

\newenvironment{jenumerate}
	{\begin{enumerate}\itemsep=-0.25em \parindent=1zw}{\end{enumerate}}
\newenvironment{jdescription}
	{\begin{description}\itemsep=-0.25em \parindent=1zw}{\end{description}}
\newenvironment{jitemize}
	{\begin{itemize}\itemsep=-0.25em \parindent=1zw}{\end{itemize}}
\renewcommand{\labelitemii}{$\cdot$}

\newcommand{\iro}[2]{{\color[named]{#1}#2\normalcolor}}
\newcommand{\irr}[1]{{\color[named]{Red}#1\normalcolor}}
\newcommand{\irg}[1]{{\color[named]{Green}#1\normalcolor}}
\newcommand{\irb}[1]{{\color[named]{Blue}#1\normalcolor}}
\newcommand{\irBl}[1]{{\color[named]{Black}#1\normalcolor}}
\newcommand{\irWh}[1]{{\color[named]{White}#1\normalcolor}}

\newcommand{\irY}[1]{{\color[named]{Yellow}#1\normalcolor}}
\newcommand{\irO}[1]{{\color[named]{Orange}#1\normalcolor}}
\newcommand{\irBr}[1]{{\color[named]{Purple}#1\normalcolor}}
\newcommand{\IrBr}[1]{{\color[named]{Purple}#1\normalcolor}}
\newcommand{\irBw}[1]{{\color[named]{Brown}#1\normalcolor}}
\newcommand{\irPk}[1]{{\color[named]{Magenta}#1\normalcolor}}
\newcommand{\irCb}[1]{{\color[named]{CadetBlue}#1\normalcolor}}

%
\newenvironment{indention}[1]{\par
\addtolength{\leftskip}{#1}\begingroup}{\endgroup\par}
%
\newcommand{\namelistlabel}[1]{\mbox{#1}\hfill} 
\newenvironment{namelist}[1]{%
\begin{list}{}
{\let\makelabel\namelistlabel
\settowidth{\labelwidth}{#1}
\setlength{\leftmargin}{1.1\labelwidth}}
}{%
\end{list}}
%
%
\newcommand{\bfig}{\begin{figure}[t]}
\newcommand{\efig}{\end{figure}}
\setcounter{page}{1}

\newtheorem{theorem}{Theorem}

\newcommand{\ep}{\mbox{\rm 2}}

\newcommand{\Exp}{\exp
}
\newcommand{\idenc}{{\varphi}_n}
\newcommand{\resenc}{
{\varphi}_n}
\newcommand{\ID}{\mbox{\scriptsize ID}}
\newcommand{\TR}{\mbox{\scriptsize TR}}
\newcommand{\Av}{\mbox{\sf E}}

\newcommand{\Vl}{|}
\newcommand{\Ag}{(R,P_{X^n}|W^n)}
\newcommand{\Agv}[1]{({#1},P_{X^n}|W^n)}
\newcommand{\Avw}[1]{({#1}|W^n)}

\newcommand{\Jd}{X^nY^n}
\newcommand{\IdR}{r_n}

\newcommand{\Index}{{n,i}}

\newcommand{\cid}{C_{\mbox{\scriptsize ID}}}
\newcommand{\cida}{C_{\mbox{{\scriptsize ID,a}}}}

\newcommand{\OMega}
{\Omega^{(\mu,\lambda,\alpha)}}
\newcommand{\ARgRv}{(p^{(n)},q_{X^n})}

\newcommand{\pmt}{\beta}

\arraycolsep 0.5mm
\date{}
%
\title{
On a Relationship between the Correct Probability of Estimation from Correlated 
Data and Mutual Information
}
\author{%
Yasutada Oohama 
\thanks{
Y. Oohama is with 
University of Electro-Communications,
1-5-1 Chofugaoka Chofu-shi, Tokyo 182-8585, Japan.
}%
}
\markboth{
}
{
}
\maketitle

\begin{abstract}
Let $X$, $Y$ be two correlated discrete random variables.
We consider an estimation of $X$ from encoded data 
$\varphi(Y)$ of $Y$ by some encoder function 
$\varphi(Y)$. We derive an inequality describing a relation
of the correct probability of estimation and the mutual 
information between $X$ and $\varphi(Y)$. This inequality 
may be useful for the secure analysis of crypto system
when we use the success probability of estimating secret 
data as a security criterion.  It also provides an intuitive 
meaning of the secrecy exponent in the strong secrecy criterion.
\end{abstract}

\section{Introduction}

It is well known that the mutual information is a very important 
quantity for an evaluation of the security of communication system. In 
the crypto system introduced by Shannon \cite{sh} perfect secrecy is 
defined by the condition that the mutual information between secret data 
and encrypted data vanishes. In the wiretap channel investigated by 
Wyner \cite{Wyn1} and in the broadcast channel with confidential 
messages investigated Csisz\'ar and K\"orner \cite{CsiKor1}, perfect 
secrecy is defined by an asymptotically vanishing {\it mutual 
information rate per channel use} between the secret messages and the 
channel outputs obtained  by the unauthorized user. 

In the several recent researches on the information theorytical security, 
the strong secrecy condition where {\it the value of mutual information} 
should asymptotically be zero is well used. Specifically, Hayashi 
\cite{HayashiIT11} has derived the relevant secrecy exponent function to 
specify the exponentially decreasing speed (i.e., exponent) of the 
leaked information under the average secrecy criterion when no cost 
constraint is considered. Han {\it et al.}\cite{HanEtAl14} extend his result to the 
case with cost constraint. The secrecy condition used by Wyner 
\cite{Wyn1} and Csisz\'ar and K\"orner \cite{CsiKor1} now called the 
weak secrecy condition has a clear intuitive meaning that the leak of 
inforamtion rate on the secret messages is asymptotically zero. On the 
other hand in the strong secrecy criterion the intuitive meaning of the 
secrecy exponent function does not seem to be so clear.

In this paper we consider a problem which is related to the intuitive 
meaning the secrecy exponent.  Our problem is as follows. Let $X$, $Y$ 
be two correlated discrete random variables. We consider an estimation 
of $X$ from encoded data $\varphi(Y)$ of $Y$ by some encoder function 
$\varphi(Y)$. We derive an inequality describing a relation of the 
correct probability of estimation and the mutual information between $X$ 
and $\varphi(Y)$. This inequality may be useful for the secure analysis 
of crypto system when we use the success probability of estimating 
secret data as a security criterion. It also provides an intuitive 
meaning of the secrecy exponent in the strong secrecy criterion.
\newcommand{\BibData}{

\bibitem{sh}C. E. Shannon, ``Communication theory of secrecy systems,''
{\em Bell Sys. Tech. Journal}, vol. 28, pp. 656-715, 1949.

\bibitem{Wyn1}A.~D.~Wyner, ``The wire-tap channel,'' 
{\em Bell Sys. Tech. Journal}, vol. 54, pp. 1355-1387, 1975.

\bibitem{CsiKor1}I.~Csisz{\' a}r and J.~K{\" o}rner, ``Broadcast
channels with confidential messages,'' 
{\em IEEE Trans. Inform. Theory}, vol. IT-24, pp. 339-348, 1978.

\bibitem{HayashiIT11} M. Hayashi,``Exponential decreasing 
rate of leaked information in universal random privacy 
amplification,'' {\em IEEE Trans. Inf. Theory}, vol. 
57, no. 6, pp. 3989-4001, 
Jun. 2011.

\bibitem{HanEtAl14}T.S. Han, H. Endo, and M. Sasaki,
``Reliability and secrecy functions of the wiretap
channel under cost constraint,'' {\em IEEE Trans. Inf. Theory}, 
vol. 60, no. 11, pp. 6819-6843, Nov. 2014.

}
\newcommand{\zap}{

Throughout in this paper, we are imposed cost constraints
(limit on available transmission energy, bandwidth, and so on).
We first address, given a general wiretap channel, the primal
problem to establish a general formula to simultaneously
summarize the reliability performance for Bob and the secrecy
performance against Eve under the maximum secrecy criterion.
Next, it is shown that both of them are described by using
exponentially decaying functions of the code length when
a stationary memoryless wiretap channel is considered. This
provides the theoretical basis for investigating the asymptotic
behavior of reliability and secrecy. We can then specifically
quantify achievable reliability exponents and achievable
secrecy exponents as well as the tradeoff between them for
several important wiretap channel models such as binary symmetric
wiretap channels, Poisson wiretap channels, Gaussian
wiretap channels. In particular, four ways of the tradeoff to
control reliability and secrecy are given and discussed with
their novel significance. Also, on the basis of the analysis of
these exponents under cost constraint, the new formula for the
ƒÂ-secrecy capacity (with the strongest secrecy among others) is
established to apply to several typical wiretap channel models.
A remarkable feature of this paper is that we first derive the
key formulas not depending on respective specific channel
models and then apply them to those respective cases to get
new insights into each case as well.
}

%
%
%
%
%

\section{Problem Statement and Results}

\subsection{Data Estimation from Correlated Data}
Let ${\cal X}$ and ${\cal Y}$ be discrete sets. We admit the case where 
those are countably infinite. Let $(X,Y)$ be a disrete random pair 
taking vaules in ${\cal X}\times {\cal Y}$ and having a probability 
distribution   
$$
p_{XY}=\left\{p_{XY}(x,y)\right\}_{(x,y)\in {\cal X} \times {\cal Y}}
$$

\begin{figure}[t]
\setlength{\unitlength}{0.94mm}

\begin{picture}(84,33)(4,0)
\put(10,27){{$X$}}
\put(10,12){{$Y$}}
\put(16,28){\vector(1,0){11}}
\put(16,13){\vector(1,0){11}}

\put(27,25){\framebox(8,6){$\phi$}}
\put(37,31){$\phi({X})$}
\put(27,10){\framebox(8,6){$\varphi$}}
\put(37,16){$\varphi(Y)$}

\put(50,28){\vector(1,-1){15}}

\put(35,28){\line(1,0){15}}
\put(35,13){\vector(1,0){30}}

\put(65,10){\framebox(7,6){$\psi$}}

\put(72,13){\vector(1,0){9}}
\put(82,12){$\hat{X}$}
\end{picture}
\vspace*{-6mm}
\caption{The case where the side information 
$\varphi(Y)$ helps an estimation of $X$ from $\phi(X)$.
}
\label{fig:Zigzag}
\noindent
\end{figure}
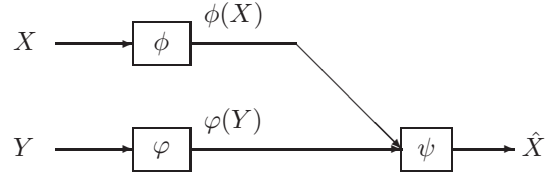

\begin{figure}[t]
\setlength{\unitlength}{0.94mm}

\begin{picture}(84,20)(4,0)
\put(10,12){{$Y$}}
\put(16,13){\vector(1,0){11}}

\put(27,10){\framebox(8,6){$\varphi$}}
\put(37,16){$\varphi(Y)$}


\put(35,13){\vector(1,0){30}}

\put(65,10){\framebox(7,6){$\psi$}}

\put(72,13){\vector(1,0){9}}
\put(82,12){$\hat{X}$}

\put(65,0){\framebox(7,6){$\psi$}}
\put(72,3){\vector(1,0){9}}
\put(82,2){$\hat{X}$}
\end{picture}
\caption{
The case where only one side information is avairable at 
the estimater and the case where no information is avairable at the estimater.
}
\label{fig:ZigzagB}
\end{figure}
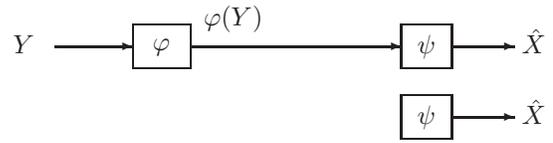

We consider a source estimation system depicted in Fig. 1. 
Data sequences $X$ and $Y$ are separately 
encoded to 
$\phi(X)$ and $\varphi(Y)$ 
and those are sent to the information processing center.
At the center the estimater $\psi$ observes 
$(\phi(X),\varphi(Y)$ to output 
the estimation $\hat{X}$ of $X$. The encoder 
functions $\phi$ and $\varphi$ are defined by
\beq
\left.
\ba{l}
\phi:{\cal X} \to {\cal M}=\left\{\,1,2,
\cdots, |{\cal M}|\right\},
\vspace{2mm}\\
\varphi:{\cal Y}\to{\cal L}
=\left\{\,1,2,\cdots, |{\cal L}|\right\}.
\ea
\right\}
\label{eqn:defen1} 
\eeq
The estimater $\psi$ is defined by 
\begin{equation}
\psi:{\cal M} \times {\cal L}\to {\cal X}.
\end{equation}
The error probability of estimation is 
\begin{equation}
{\rm P}_{\rm e}(\phi,\varphi,\psi| p_{XY})
=\Pr\left\{\hat{X} \neq X\right\}, 
\end{equation}
where $\hat{X}=\psi(\phi(X),\varphi(Y))$.
The correct probability of estimation is 
\begin{equation}
{\rm P}_{\rm c}(\phi,\varphi,\psi| p_{XY})
=1-{\rm P}_{\rm e}(\phi,\varphi,\psi| p_{XY})
=\Pr\left\{ \hat{X} = X\right\}. 
\end{equation}
We condsier the following three cases. 
\begin{itemize}
\item[1.]The case where the side information 
$\varphi(Y)$ serves as a helper to estimate $X$ from $\phi(X)$.(Case 1) 
\item[2.]The case where only the helper $\varphi(Y)$ 
is avairable for an estimation of $X$.(Case 2) 
Case 2 corresponds to the case where $|{\cal M}|=1$ and 
$\phi$ is a constant function given by $\phi(x)=1,x \in{\cal X}$. 
The decoder function $\psi$ in this case is given by 
$\psi:{\cal L}\to{\cal X}$.
\item[3.]The case where no information 
is avairable for an estimation of $X$.(Case 3) 
Case 3 corresponds to the case where 
$|{\cal M}|=|{\cal L}|=1$ and $\phi$ and $\varphi$ are constant 
functions given by $\phi(x)=1, x\in{\cal X}$ and 
$\varphi(y)=1, y\in{\cal Y}$. 
The decoer function $\psi$ in this case is given by 
$\psi:\{ 1 \}\to{\cal X}$.
\end{itemize}
Let the correct probability of estimation in Case 2 is denoted by 
${\rm P}_{\rm c}(\varphi,\psi|p_{XY}).$
Let the correct probability of estimation in Case 3 is denoted by 
${\rm P}_{\rm c}(\psi|p_{XY}).$
Set

\beqno
{\rm P}_{\rm c, \max}^{(1)}(p_{XY})
&\defeq & \max_{\scs \phi: {\cal X}\to {\cal M},
   \atop{\scs \varphi: {\cal Y} \to {\cal L},
       \atop{ \scs \psi: {\cal M} \times {\cal L} \to {\cal X}
       }
   }
}{\rm P}_{\rm c}(\varphi,\psi|p_{XY}), 
\\
{\rm P}_{\rm c, \max}^{(2)}(p_{XY})
&\defeq & \max_{\scs 
\varphi: {\cal Y}\to {\cal L},
\atop{\scs \psi: {\cal L} \to {\cal X}}
        }{\rm P}_{\rm c}(\varphi,\psi|p_{XY}), 
\\
{\rm P}_{\rm c, \max}^{(3)}(p_{XY})
& \defeq & \max_{\psi: \{1\}\to {\cal X}}{\rm P}_{\rm c}(\psi|p_{XY}). 
\eeqno
Our aim is to clarify relationships  between the above three quantities.
By definition it is obvious that 
$$
{\rm P}_{\rm c, \max}^{(1)}(p_{XY})\geq 
{\rm P}_{\rm c, \max}^{(2)}(p_{XY})\geq 
{\rm P}_{\rm c, \max}^{(3)}(p_{XY}).
$$
Set 
$$
p_{\max} \defeq \max_{x\in {\cal X}}p_{X}(x).
$$
Then we have  
\beqno
{\rm P}_{\rm c, \max}^{(3)}(p_{XY})
&=&\max_{\scs \psi: \{1\} \to {\cal X},
 \atop{\scs x \in {\cal X}:\psi(1)=x}} p_X(x)
\\
&=&\max_{\psi(1) \in {\cal X}}p_{X}(\psi(1))=p_{\max}.
\eeqno
We are particularly interested in a difference between
${\rm P}_{\rm c, \max}^{(2)}($ $p_{XY})$ and 
${\rm P}_{\rm c, \max}^{(3)}(p_{XY})$. If there is no 
difference between those to quantities. The side information 
$\varphi(Y)$ is of no use to estimate $X$. 
In this paper we derive an inequality stating that 
the difference is upper bounded by the mutual information between 
the side information $\varphi(Y)$ and the source $X$.

\subsection{Main Results}
In this subsection we sate our main result. 
We first give a proposition 
which plays a key role in deriving our main results.   
Set $S=\varphi(Y)$. The joint 
distribution $p_{XY}$ of $(X,Y,S)$ is given by 
\beqno
p_{XYS}(s,x,y)=p_{XY}(x,y)p_{S|Y}(s|y). 
\eeqno
It is obvious that the random variables $X,Y,S$ form Markov chain 
$X \markov Y \markov S$. The following proposition 
providing an upper bound of ${\rm P}_{\rm c, \max}^{(0)}(p_{XY})$ 
is useful to derive our main result. 

\begin{pro}\label{pro:OhzzzPP}
For any $\eta>0$ and for any $(\phi$, $\varphi,\psi)$, 
we have 
\beqno
& &{\rm P}_{\rm c}(\phi,\varphi,\psi | p_{XY})
\nonumber\\
&\leq &
p_{SX}\left\{\ds 
\log |{\cal M}| \geq  \log\frac{1}{ p_{X|S}(X|S) }-\eta
\right\}
+\ep^{-\eta}. 
\label{eqn:azsadZZ}
\eeqno
Specifically, we have
\beqno
& &{\rm P}_{\rm c, \max}^{(1)}(p_{XY})
\nonumber\\
&\leq &
p_{SX}\left\{\ds 
\log |{\cal M}| \geq  \log\frac{1}{ p_{X|S}(X|S) }-\eta
\right\}
+\ep^{-\eta}. 
\eeqno
\end{pro}

Proof of this proposition is given in the next section. 
Using this proposition, we obtain the following result. 
\begin{Th}\label{Th:mainThss}
For any $\nu\in (0,\log\frac{1}{p_{\max}})$, we have 
\beq
{\rm P}_{\rm c, \max}^{(2)}(p_{XY})\leq 
\ep^{\nu}(p_{\max})+ \frac{1}{\nu}I(X;\varphi(Y)).
\label{eqn:Aww}
\eeq
\end{Th}

Proof of this theorem is given in the next section. 
From Theorem \ref{Th:mainThss} and 
$
\ep^{\nu}\leq 1+\nu \mbox{ for } \nu \in [0,1],
$
we have the following corollary.
\begin{co}
For any $\nu \in (0, \min\{1, \log \frac{1}{p_{\max}} \})$, we have 
\beqno
{\rm P}_{\rm c, \max}^{(2)}(p_{XY})\leq 
(1+\nu)p_{\max} + \frac{1}{\nu}I(X;\varphi(Y)).
\eeqno
\end{co}

\section{Proofs of the Results} 

In this section we prove Proposition \ref{pro:OhzzzPP} and Theorem 
\ref{Th:mainThss}. We first prove Proposition \ref{pro:OhzzzPP}. To 
prove this proposition, we prepare a lemma. Set
\beqno
& &{\cal D} \defeq
\{(s,x,y): s=\varphi(y), p_{X|S}(x|s)\geq (1/|{\cal M}|)\ep^{-\eta}\},
\\
&&{\cal E}
\defeq \{ (s,x,y): s = \varphi(y),\psi(\phi(x),\varphi(y))=x \}.
\eeqno
Then we have the following lemma. 
\begin{lm}\label{lm:zzxaZZ}
\beqno
p_{SXY}
\left({\cal D}^{\rm c}\cap {\cal E} \right) \leq \ep^{-\eta}.
\eeqno
\end{lm}

{\it Proof:} 
We first observe that 
\beqno
& &p_S(s)=\sum_{\scs y: \varphi(y)=s}p_{Y}(y),\:
p_{Y|S}(y|s)=\frac{p_{Y}(y)}{p_{S}(s)}.
\eeqno
We have the following chain of inequalities:
\beqno
& &p_{SXY}
\left({\cal D}^{\rm c}\cap {\cal E}\right)
\\
&=&\sum_{s \in{\cal L}}p_S(s)
\sum_{\scs y: \varphi(y)=s}p_{Y|S}(y|s)
\\
& &
\times
\sum_{\scs x: \psi(s,\phi(x))=x  
      \atop{\scs
           p_{X|S}(x|s) < (1/|{\cal M}|)\ep^{-\eta}
     }
}
p_{X|Y}(x|y)
\\
&=&\sum_{\scs s\in{\cal L}}p_S(s)
\sum_{\scs x:  \psi(s,\phi(x))=x
      \atop{\scs
           p_{X|S}(x|s) < (1/|{\cal M}|)\ep^{-\eta}
     }
}
p_{X|S}(x|s)
\\
&\leq &
\sum_{s\in{\cal L}}p_S(s)
\frac{1}{|{\cal M}| }\ep^{-\eta}
\left|\left\{x:\psi(s,\phi(x))=x
\right\}\right|
\\
&\MLeq{a}&\sum_{s\in{\cal L}}p_S(s) 
\frac{1}{|{\cal M}|}\ep^{-\eta}|{\cal M}|
=\ep^{-\eta}.
\eeqno
Step (a) follows from that the number of 
$x \in {\cal X}$ correctly decoded does not exceed $|{\cal M}|$.   
\hfill\IEEEQED

{\it Proof of Proposition \ref{pro:OhzzzPP}:} By definition we have
\beqno
& &p_{SXY}\left({\cal D}\right)=
p_{SX}
\left\{
\log |{\cal M}| \geq \log \frac{1}{p_{X|S}(X|S)}-\eta
\right\}.
\eeqno
Hence, it suffices to show 
\beqno
{\rm P}_{\rm c}(\phi,\varphi,\psi|p_{XY})
&\leq&
p_{SXY}\left({\cal D}\right) +\ep^{-\eta}
\eeqno
to prove Proposition \ref{pro:OhzzzPP}. By definition 
we have
\beqno 
&&{\rm P}_{\rm c}(\phi,\varphi,\psi|p_{XY})=p_{SXY}\left({\cal E}\right).
\eeqno
Then we have the following.
\beqno
&&{\rm P}_{\rm c}(\phi,\varphi,\psi|p_{XY})=p_{SXY}\left({\cal E}\right)
\\
&=& p_{SXY}\left({\cal D} \cap {\cal E}\right)
+p_{SXY}\left({\cal D}^{\rm c}\cap {\cal E}\right)
\\
&\leq&
 p_{SXY}\left({\cal D}\right)
+p_{SXY}\left({\cal D}^{\rm c}\cap {\cal E} 
\right)
\MLeq{a}
p_{SXY}\left( {\cal D}\right)+\ep^{-\eta}. 
\eeqno
Step (a) follows from Lemma \ref{lm:zzxaZZ}.
\hfill\IEEEQED

{\it Proof of Theorem \ref{Th:mainThss}: } We have the following chain 
of inequalities:
 
\begin{align}
& {\rm P}_{\rm c, \max}^{(1)}(p_{XY})
\MLeq{a} 
p_{SX}\left\{\ds 
0 \geq \log\frac{1}{ p_{X|S}(X|S) }-\eta
\right\}
+\ep^{-\eta} 
\nonumber\\
&=p_{SX}\left\{
\left[\ds \log\frac{1}{p_{X|S}(X|S)}\leq \eta\right]\right.
\nonumber\\
& \qquad\qquad \qquad \left. \bigcap\left[\log \frac{p_{X|S}(X|S)}{p_{X}(X)}<\nu
\right]
\right\}
\nonumber\\
&\quad +p_{SX}\left\{
      \left[ \log\frac{1}{p_{X|S}(X|S)}\leq \eta\right]
      \right.
\nonumber\\
&\qquad\qquad\qquad  \left. 
\bigcap \left[\log \frac{p_{X|S}(X|S)}{p_{X}(X)} \geq \nu\right]
       \right\}
+\ep^{-\eta} 
\nonumber\\
&\leq
 p_{X}\left\{\ds \log\frac{1}{p_{X}(X)}< \eta+\nu \right\}
\nonumber\\
&\quad +p_{SX}\left\{ \log \frac{p_{X|S}(X|S)}{p_{X}(X)} \geq \nu\right\}
+\ep^{-\eta} 
\nonumber\\
&\MLeq{b}
 p_{X}\left\{
\ds \log\frac{1}{p_{X}(X)}< \eta+\nu \right\}
\nonumber\\
&\quad +\frac{1}{\nu}{\rm E}_{p_{SX}}
\left[\log \frac{p_{X|S}(X|S)}{p_{X}(X)}\right]
+\ep^{-\eta} 
\nonumber\\
&=p_{X}\left\{
\ds \log\frac{1}{p_{X}(X)}< \eta+\nu \right\}
+\frac{1}{\nu}I(X;S)
+\ep^{-\eta}.
\label{eqn:sdEe} 
\end{align}
In step (a) we use Proposition \ref{pro:OhzzzPP} for $|{\cal M}|=1$.  
Step (b) follows from the Markov's inequality. 
In (\ref{eqn:sdEe}), we choose $\eta,\nu$ so that
$$
\eta+\nu= \log \frac{1}{p_{\max}}=
\min_{x \in {\cal X}}\log \frac{1}{p_X(x)}.  
$$
Since 
$$
\eta= -\nu + \log \frac{1}{p_{\max}}>0,
$$
$\nu$ must satisfy $\nu \in (0, \log \frac{1}{p_{\max}})$,
For this chooice of $\eta,\nu$, we have 
\beq
p_{X}\left\{\ds \log\frac{1}{p_{X}(X)}< \eta+\nu \right\}
=0,
\ep^{-\eta}=\ep^{\nu}(p_{\max}).
\label{eqn:paa}
\eeq
From (\ref{eqn:sdEe}) and (\ref{eqn:paa}), 
we have the bound (\ref{eqn:Aww}) in Theorem \ref{Th:mainThss}. 
\hfill\IEEEQED

\end{document}